\documentclass[conference]{IEEEtran}
\IEEEoverridecommandlockouts
\usepackage{amsmath,amssymb,amsfonts}
\usepackage{algorithmic}
\usepackage{graphicx}
\usepackage{textcomp}
\usepackage{xcolor}
\usepackage[numbers]{natbib}
\usepackage{personalcommand}
\usepackage{cleveref}

\def\BibTeX{{\rm B\kern-.05em{\sc i\kern-.025em b}\kern-.08em
    T\kern-.1667em\lower.7ex\hbox{E}\kern-.125emX}}
\begin{document}

\title{Multi-objective Software Architecture Refactoring driven by Quality Attributes}

\author{\IEEEauthorblockN{Daniele Di Pompeo}
\IEEEauthorblockA{\textit{University of L'Aquila} \\
L'Aquila, Italy \\
daniele.dipompeo@univaq.it}
\and
\IEEEauthorblockN{Michele Tucci}
\IEEEauthorblockA{\textit{Charles University} \\
Prague, Czech Republic \\
tucci@d3s.mff.cuni.cz}
}

\maketitle

\begin{abstract}
Architecture optimization is the process of automatically generating design options, typically to enhance software's quantifiable quality attributes, such as performance and reliability.
Multi-objective optimization approaches have been used in this situation to assist the designer in selecting appropriate trade-offs between a number of non-functional features.
Through automated refactoring, design alternatives can be produced in this process, and assessed using non-functional models. 

This type of optimization tasks are hard and time- and resource-intensive, which frequently hampers their use in software engineering procedures.

In this paper, we present our optimization framework where we examined the performance of various genetic algorithms.
We also exercised our framework with two case studies with various levels of size, complexity, and domain served as our test subjects. 

\end{abstract}


\section{Introduction}\label{sec:intro}

Different factors, such as the addition of new requirements, the adaption to new execution contexts, or the deterioration of non-functional features, can lead to software refactoring.
The challenge of identifying the best refactoring operations is challenging because there is a wide range of potential solutions and no automated assistance is currently available.

In this situation, search-based approaches have been widely used \citep{Mariani:2017jd,Ouni:2017db,Ramirez:2018uz,Ray:2014ip,Aleti:2013gp}.

Multi-objective optimization approaches, which are search-based, have lately been used to solve model refactoring optimization issues~\citep{CORTELLESSA2021106568,NI2021106565}.
Searching among design alternatives (for example, through architectural tactics) is a typical feature of multi-objective optimization methodologies used to solve model-based software restructuring challenges~\citep{Koziolek:2011cg,NI2021106565}. 

In this study, we describe a many-objective evolutionary framework that automatically searches and applies sequences of refactoring actions leading to the optimization of four objectives: i) performance variation (analyzed through Layered Queueing Networks~\citep{DBLP:journals/tse/NeilsonWPM95}), ii) reliability (analyzed through a closed-form model~\citep{CortellessaSC02}), iii) number of performance antipatterns (automatically detected~\citep{DBLP:journals/infsof/ArcelliCP18}), and iv) architectural distance~\citep{Arcelli:2018vo}.
 
In particular, our framework automatically applies refactoring actions to the initial architecture, and we analyze the contribution of the architectural distance to the generation of Pareto frontiers~\citep{DBLP:conf/wcre/ArcelliCP19}.
Furthermore, we study the impact of performance antipatterns on the quality of refactoring solutions. 
Since it has been shown that removing performance antipatterns leads to systems that show better performance than the ones affected by them~\cite{DBLP:journals/infsof/ArcelliCP18}, we aim at studying if this result persists in the context of many-objective optimization, where performance improvement is not the only objective. 

Our approach applies to UML augmented by MARTE \cite{MARTE} and DAM \cite{BernardiMP11} profiles that allow to embed performance and reliability properties. However, UML does not provide native support for performance analysis, thus we introduce a model-to-model transformation that generates Layered Queueing Networks (LQN) from annotated UML artifacts.
The solution of LQN models feeds the performance variation objective.

Here, we consider refactoring actions that are designed to improve performance in most cases~\cite{DBLP:conf/icsa/ArcelliCPET19,DBLP:journals/jss/CortellessaPET22}.
Since such actions may also have an impact on other non-functional properties, we introduce the reliability among the optimization objectives to study whether satisfactory levels of performance and reliability can be kept at the same time.
In order to quantify the reliability objective, we adopt an existing model for component-based software systems~\cite{CortellessaSC02} that can be generated from UML.

We also minimize the distance between the initial architecture and the ones resulting from applying refactoring actions. Indeed, without an objective that minimizes such distance, the proposed solutions could be impractical because they could require to completely disassemble and re-assemble the initial architecture. 

In a recent work~\cite{SEAA2021}, we extended the approach in~\cite{Arcelli:2018vo,CORTELLESSA2021106568}, by investigating architecture optimization, thus widening the scope of eligible models. 
We analyze the sensitivity of the search process to configuration variations. We refine the cost model of refactoring actions and we investigate how it contributes to the generation of Pareto frontiers. 

The experimentation lasted several hours and generated thousands of model alternatives.
Generally, multi-objective optimization is beneficial when the solution space is so large that an exhaustive search is impractical. 
Hence, due to the search of the solution space, multi-objective optimization requires a lot of time and resources.

Finally, to encourage reproducibility, we publicly share the implementation of the approach~\footnote{\url{https://github.com/SEALABQualityGroup/EASIER}}, as well as the data gathered during the experimentation~\footnote{\url{https://github.com/SEALABQualityGroup/2022-ist-replication-package}}.
\section{Related Work}\label{sec:related}

In the past ten years, studies on software architecture multi-objective optimization have been developed to optimize various quality attributes (such as reliability and energy)~\cite{Martens:2010bn,5949650,DBLP:conf/qosa/MeedeniyaBAG10,10.1007/978-3-642-13821-8_8,CORTELLESSA2021106568}; with various degrees of freedom in the modification of architectures (such as service selection~\cite{Cardellini:2009:QRA:1595696.1595718}.

Recent research analyzes the capacity of two distinct multi-objective optimization algorithms to enhance non-functional features inside a particular architecture notation (i.e., Palladio Component Model)~\cite{NI2021106565,10.1145/3132498.3132509,Becker:2009cl}.
The authors use architectural approaches to find the best solutions, which primarily include changing system parameters (such as hardware settings or operation requirements).
On the other hand, in this work, we employ refactoring techniques that alter the basic architecture structure while keeping the original behavior.
The architecture notation is another difference; rather than using a unique Domain Specific Language, we use UML with the intention of experimenting with a standard notation.

Menasce et al. have provided a framework for architectural design and quality optimization, \cite{DBLP:conf/wosp/MenasceEGMS10}. This framework makes use of architectural patterns to help the search process (such as load balancing and fault tolerance).
The approach has two drawbacks: performance indices are computed using equation-based analytical models, which may be too simple to capture architectural details and resource contention; the architecture must be designed in a tool-specific notation rather than in a standard modeling language (as we do in this paper). 

A method for modeling and analyzing AADL architectures has been given by Aleti et al.\cite{DBLP:books/daglib/0030032}.
A tool that may be used to optimize various quality attributes while adjusting architecture deployment and component redundancy has also been introduced.
Our framework, instead, makes use of UML and takes into account more intricate refactoring procedures as well as various goal attributes for the fitness function.
In addition, we look into the function of performance antipatterns in the context of optimizing many-objective architecture refactoring. 

\section{Approach}\label{sec:approach}

\begin{figure}
      \centering
      \includegraphics[width=0.97\linewidth]{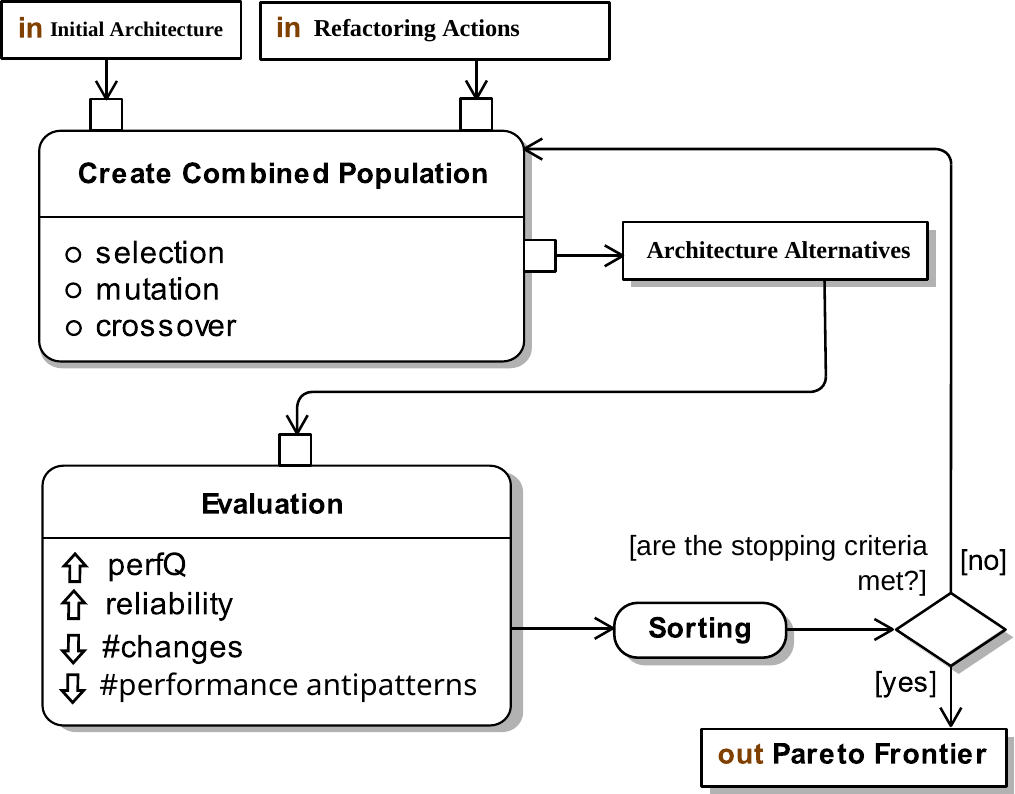}
      \caption{Our multi-objective evolutionary approach}
      \label{fig:process}
\end{figure}

The process that we describe in this research is illustrated in \Cref{fig:process}.

An \emph{Initial Architecture} and a list of refactoring actions are supplied into the process.
The \emph{Create Combined Population} step, where mating operations (\ie selection, mutation, and crossover) are implemented to create \emph{Architecture Alternatives} involves the \emph{Initial Architecture} and the \emph{Refactoring Actions}.
The refactoring activities are randomly and automatically applied by the mating operations, producing alternatives that are functionally comparable to the initial architecture.

Therefore, each architecture alternative is given the \emph{Evaluation step}.
The model options are then sorted (\emph{Sorting step}) based on the following four goals: \emph{\perfq, \reliability, \achanges, and \pas}.
Throughout the process, these qualities are appraised and taken into consideration to select the optimal candidates. 

Recently, we investigated how performance antipatterns affect the effectiveness of refactoring methods~\citep{SEAA2021}.
We aim to investigate whether this phenomenon also holds in the context of multi-objective optimization, where performance improvement is not the only goal, given that it has been demonstrated that removing performance antipatterns results in systems that show better performance than those affected by them~\citep{DBLP:conf/cmg/SmithW01a,Smith:2003wv,DBLP:journals/infsof/ArcelliCP18}. 

Furthermore, we looked into whether adding a time budget could shorten the amount of time an evolutionary algorithm requires~\citep{SEAA2022}.
The purpose of setting such a time constraint is to determine the extent to which, in a model-based multi-objective refactoring optimization scenario, the imposition of a time-based search budget can degrade the quality of the resultant Pareto fronts.
Furthermore, we are curious about how various algorithms respond to various search budgets.
In order to test this, we chose two case studies and ran the optimization with search budgets of \emph{15}, \emph{30}, and \emph{60} minutes. 

Currently, our framework supports three genetic algorithms, \nsga~\citep{Deb:2002ut}, \spea~\citep{zitzler2001spea2}, \pesa~\citep{Corne_Jerram_Knowles_Oates_2001}. We selected these algorithms with respect their different searching policies. Thus, our results cover evolutionary algorithms of different characteristics.

\section{Conclusion and Future Work}\label{sec:conclusion}

We have developed a framework for multi-objective architecture optimization that takes into account quality attributes.
In the context of architecture optimization, we concentrated our investigation on the potential effects of evolutionary algorithms on the quality of optimal refactoring solutions. 

We learned some interesting things from our experimentation concerning the effectiveness of the created solutions and the use of performance antipatterns as an algorithmic objective.
In this regard, we demonstrated that we may achieve superior solutions in terms of performance and reliability by incorporating the detection of performance antipatterns into the optimization process.
Making sure that our strategy did not decrease the reliability of the basic architecture was another crucial component of our investigation.
Our tests revealed that, in most instances, we were able to boost the reliability of alternatives in comparison to the original architecture. 

Future research will examine how settings (experiment and algorithm setups) affect the effectiveness of Pareto frontiers.
We will examine the effects of denser populations, for instance, on calculation time and the accuracy of computed Pareto frontiers.
Our research focuses on the impact of predicting the baseline refactoring factor using a more complex cost model, such as COCOMO-II~\citep{boehm2009software}, on the combination of refactoring activities. We are also interested in the influence that changes play.
We want to expand the portfolio of refactoring activities, for instance by adding fault tolerance refactoring actions~\cite{DBLP:journals/infsof/CortellessaET20}, and a fruitful inquiry will focus on the length of the sequence of refactoring actions, which is presently fixed to four refactoring actions. 
We will incorporate additional evolutionary algorithms into our approach to examine the role that various optimization methods play in the architecture refactoring.

\section*{Acknowledgment}
Daniele Di Pompeo is supported by the Centre of EXcellence on Connected, Geo-Localized and Cybersecure Vehicle (EX-Emerge), funded by the Italian Government under CIPE resolution n. 70/2017 (Aug. 7, 2017).

Michele Tucci is supported by the OP RDE project No. CZ.02.2.69/\-0.0/\-0.0/\-18\_053/\-0016976 ``International mobility of research, technical and administrative staff at the Charles University''.

\bibliographystyle{IEEEtran}
\bibliography{IEEEabrv,biblio}

\end{document}